\documentclass[prb,twocolumn,notitlepage,showpacs, superscriptaddress,aps,amsmath,amssymb,UFT8]{revtex4-1}
\usepackage[usenames,dvipsnames]{color} 
\usepackage{graphicx}
\usepackage{bm}
\usepackage{CJK}
\definecolor{LinkColor}{rgb}{0,0,.5}
\usepackage[colorlinks=true,linkcolor=BrickRed,citecolor=LinkColor,urlcolor=LinkColor]{hyperref}
\newcommand{\carb}{$^{13}$C }
\newcommand{\Nit}{$^{14}$N }
\newcommand{\ket}[1]{\left\vert{#1}\right\rangle}
\newcommand{\bra}[1]{\left\langle{#1}\right\vert}

\newcommand{\ave}[1]{\left\langle #1\right\rangle}
\newcommand{\ham}{{\mathcal{H}}}

\newcommand{\sy}{\sigma_y}

\begin{document}
\begin{CJK*}{UTF8}{} 

\title{Measurement of transverse hyperfine interaction by forbidden transitions}

\author{Mo Chen \CJKfamily{gbsn}(陈墨) }
\thanks{These authors contributed equally to this work.}
\affiliation{
Research Laboratory of Electronics, Massachusetts Institute of Technology, Cambridge, Massachusetts 02139, USA
}
\affiliation{
Department of Mechanical Engineering, Massachusetts Institute of Technology, Cambridge, Massachusetts 02139, USA
}
\author{Masashi Hirose}
\thanks{These authors contributed equally to this work.}
\affiliation{
Research Laboratory of Electronics, Massachusetts Institute of Technology, Cambridge, Massachusetts 02139, USA
}
\affiliation{
Department of Nuclear Science and Engineering, Massachusetts Institute of Technology, Cambridge, Massachusetts 02139, USA
}
\author{Paola Cappellaro }
\thanks{pcappell@mit.edu}
\affiliation{
Research Laboratory of Electronics, Massachusetts Institute of Technology, Cambridge, Massachusetts 02139, USA
}
\affiliation{
Department of Nuclear Science and Engineering, Massachusetts Institute of Technology, Cambridge, Massachusetts 02139, USA
}

\begin{abstract}
Precise characterization of a system's Hamiltonian is crucial to its high-fidelity control that would enable many quantum technologies, ranging from quantum computation to communication and sensing. In particular, non-secular parts of the Hamiltonian are usually more difficult to characterize, even if they can give rise to subtle but non-negligible effects. Here we present a strategy for the precise estimation of the transverse hyperfine coupling between an electronic and a nuclear spin, exploiting effects due to nominally forbidden transitions during the Rabi nutation of the nuclear spin. We applied the method to precisely determine the transverse coupling between a Nitrogen-Vacancy center electronic spin and its Nitrogen nuclear spin. In addition, we show how this transverse hyperfine coupling, that has been often neglected in experiments, is crucial to achieving large enhancements of the nuclear Rabi nutation rate. 
\end{abstract}

\maketitle
\end{CJK*}
Quantum technologies promise to revolutionize many fields, ranging from precision sensing to fast computation. The success of novel technologies based on quantum effects rests on engineering quantum systems robust to noise and decoherence and on controlling them with high precision. Solid-state systems comprising nuclear spins have emerged as promising candidates, since the nuclear spin qubits are only weakly coupled to external fields and thus exhibit long coherence times. In order for nuclear spins to be used as good qubits, there are two important requirements: their Hamiltonians need to be known with very high precision, as this would enable applying e.g. optimal control methods\cite{Khaneja05,Scheuer14}, and strong driving should be available, in order to achieve fast gates. Here we show how to meet these two requirements by exploiting nominally forbidden transitions in a hybrid electronic-nuclear spin system associated with the Nitrogen-Vacancy center in diamond\cite{Gruber97}. Specifically, we use second-order effects due to mixing of the electronic and nuclear spin states\cite{AbragamBleaney} in order to identify with high precision their coupling strength and to enhance the nuclear spin nutation rate\cite{Sangtawesin15}.

The nitrogen vacancy (NV) center is a naturally occurring point defect in diamond\cite{Field92}. Thanks to its optical properties and long coherence times, it has emerged as a versatile system for quantum sensing\cite{Taylor08,Dolde11,Toyli13}, quantum information\cite{Wrachtrup01,Cappellaro09} and photonics applications\cite{Aharonovich11,Hausmann12}. The nuclear \Nit spin often plays an important role in these applications. Not only can it serve as a qubit in small quantum algorithms\cite{Fuchs11,Waldherr11,George13}, but it can also be used to enhance the readout fidelity of the NV electronic spin\cite{Neumann10b} and achieve more sensitive detection of magnetic fields\cite{Kessler14,Arrad14} and rotations\cite{Ledbetter12,Ajoy12g}. These applications are made possible by the hyperfine interaction between the NV electronic and nuclear spins.

While the secular part of the NV-\Nit Hamiltonian has been well-characterized before\cite{Steiner09,Smeltzer09,Shin14}, the transverse hyperfine coupling is more difficult to measure\cite{Kempf00} and published values do not match well\cite{Doherty13,Felton09,He93b}. The most precise characterization to date has been  achieved by ensemble ESR techniques\cite{Felton09}. 
In that work, the ESR spectrum of an ensemble of NV centers was measured by induction methods while applying a magnetic field along the $\ave{110}$ direction to amplify nominally forbidden transitions. This method is not applicable to single NV centers, since the strong transverse field would quench the spin-dependent optical contrast. 

Here we propose a different strategy to measure the transverse hyperfine coupling that can be carried out with optically detected magnetic resonance.  Thanks to this method we can determine the value of the transverse coupling with a better precision than achieved previously. The method is not restricted to the NV spin system, but could be applied more generally to other electronic-nuclear spin systems, such as phosphorus\cite{Morton08} or antimony\cite{Wolfowicz14} donors in silicon, defects in silicon carbide\cite{Widmann15,Falk15} or quantum dots\cite{Chekhovich13}. 
Precise knowledge of the hyperfine interaction tensor would enable achieving more precise control, elucidating modulations of the NV echo dynamics
or, as we show here, achieving faster Rabi nutation of the nuclear spin. 
\begin{figure*}\begin{center}
\includegraphics[width= 0.4\textwidth]{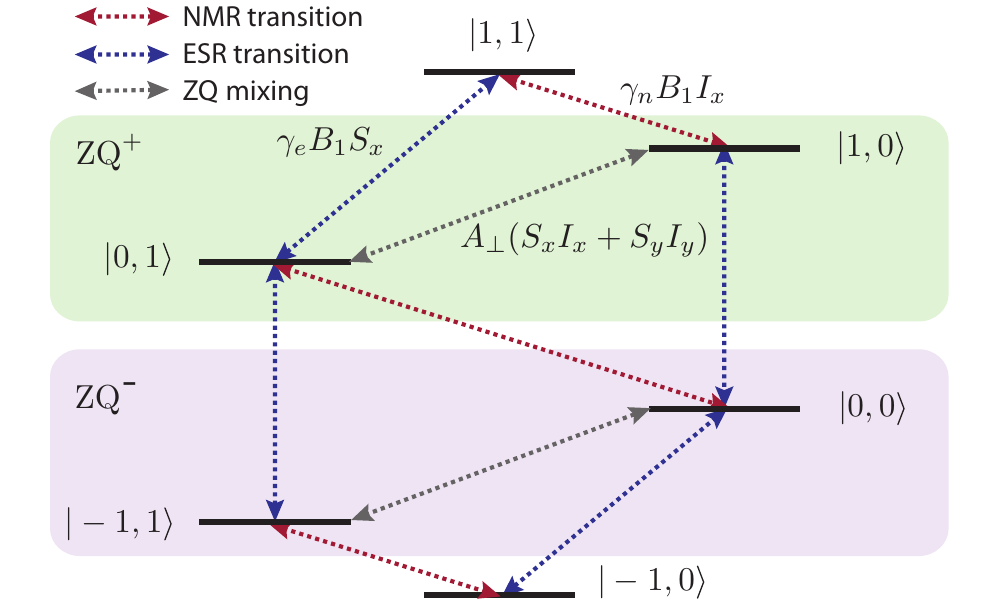}\qquad\qquad\includegraphics[width= 0.4\textwidth]{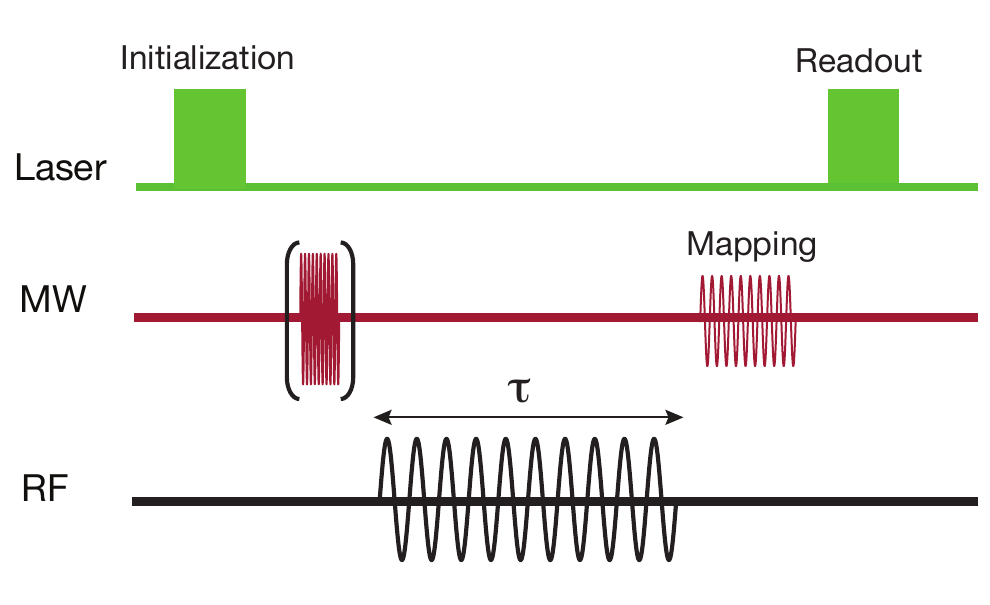}\end{center}
\caption{Left: energy levels of the reduced NV-\Nit spin system, showing the transitions that are mixed by the transverse hyperfine coupling. Right: 
Experimental sequence used to measure the nuclear \Nit Rabi frequency in the three NV manifolds.\label{fig:experiment}}
\end{figure*}

\textbf{Theoretical Model --}
The NV ground state is a two-spin system given by the electronic spin of the NV center ($S = 1$) and the nuclear spin ($I = 1$) of the substitutional \Nit adjacent to the vacancy that comprise the defect. In the experiments, we are only interested in two of the nuclear spin levels ($m_I=+1,0$) that we drive on-resonance, while the third level can be neglected. Then, the Hamiltonian of the reduced system\cite{SMAPS} is given by $\mathcal{H}=\mathcal{H}_\parallel+\mathcal{H}_\perp$, where the secular, $\mathcal{H}_\parallel$, and nonsecular, $\mathcal{H}_{\perp}$, terms are: 
\begin{eqnarray}
\mathcal{H}_\parallel & = & \Delta S_z^2+(\gamma_eB_z+\frac{A_\parallel}{2})  S_z +(Q+\gamma_nB_z)I_z 
					+A_\parallel S_zI_z, \nonumber		\\				
\mathcal{H}_\perp &=&\sqrt{2}A_\perp(S_xI_x+S_yI_y).
\label{eq:NVHam1}
\end{eqnarray}
Here $S$ and $I$ are the electron spin-1  and nuclear spin-1/2 operator respectively. $\Delta=2.87$~GHz is the zero-field splitting and $Q = -4.945$~MHz\cite{Smeltzer09} the nuclear quadrupolar interaction. The NV spin is coupled to the nuclear spin by a hyperfine interaction  with a longitudinal component $A_\parallel = -2.162$~MHz\cite{Smeltzer09} and a transverse component $A_\perp$ which we want to estimate. A magnetic field $B_z$ is applied along the NV crystal axis [111] to lift the degeneracy of the $m_s=\pm1$ level, yielding the electron and nuclear Zeeman frequencies $\gamma_eB_z$ and $\gamma_nB_z$ where $\gamma_e = 2.8$~MHz/G and $\gamma_n = - 0.308$ kHz/G. 

Let $\ket{m_s,m_I}$ be  eigenstates of $\mathcal{H}_\parallel$. The transverse coupling $A_\perp$ mixes states connected via zero-quantum (ZQ) transitions, $\ket{+1,0}\leftrightarrow \ket{0,1}$ and $\ket{0,0} \leftrightarrow \ket{\text{-}1,1}$. Diagonalization of the total Hamiltonian can then be achieved by rotating the two ZQ subspaces with  a unitary transformation  $U_{\rm ZQ}=e^{-i(\sy^-\vartheta^-+\sy^+\vartheta^+)}$, where we defined $\sy^+=i(\ket{+1,0}\!\bra{0,1}-\ket{0,1}\!\bra{+1,0})$;  $\sy^-=i(\ket{0,0}\!\bra{\text{-}1,1}-\ket{\text{-}1,1}\!\bra{0,0})$ and the rotation angles are
\vspace{-10pt}
\begin{equation}\renewcommand\arraystretch{2.5}
\begin{array}{lcl}
\tan (2 \vartheta^+) & = & \displaystyle \frac{2A_\perp}{\Delta+\gamma_eB_z-\gamma_nB_z-Q} ,\\
\tan (2 \vartheta^-) & = & \displaystyle\frac{-2A_\perp}{\Delta-\gamma_eB_z-A_{\parallel}+\gamma_nB_z+Q}.
\end{array}
\label{eq:DiagAngle}	
\end{equation}
Because of this level mixing, a field on resonance with the nuclear spin transition also drives electronic transitions. Although the electronic spin state is unchanged to first order, as long as the mixing is small, the nominally forbidden transitions result in an enhancement of the nuclear state nutation frequency, as we explain below.

	When applying a radio frequency (RF) field to drive the nuclear spin, the interaction Hamiltonian of the NV-\Nit system with the RF field is:	
	\begin{equation}
	\mathcal{H}_{\rm rf}(t) =2 B_1 \cos(\omega t) (\gamma_e	 S_x +\sqrt{2}\gamma_n I_x),
	\label{eq:RFHam}
	\end{equation}
	where $B_1$ is the RF field strength. The Hamiltonian can be simplified by going into a rotating picture at the RF frequency $\omega$ and applying the rotating wave approximation (RWA), to obtain $\mathcal{H}_{\rm rf} = B_1  (\gamma_e	 S_x +\sqrt{2}\gamma_n I_x)$. 
	We note that since we might have $\gamma_eB_1\gg \omega$, effects from the counter-rotating fields, such as Bloch-Siegert shifts of the electronic energies, might be present. These effects were however negligible at the fields and Rabi strengths used in the experiments\cite{SMAPS}. Transforming  $\mathcal{H}_{\rm rf}$ with the unitary $U_{\rm ZQ}$ and denoting states and operators in the new frame by a hat, we obtain $\hat{\mathcal{H}}_{\rm rf}\!=\!U_{\rm ZQ}\mathcal{H}_{\rm rf}(t)U^\dagger_{\rm ZQ}\!=\!\ham_n+\ham_e$, with
\begin{equation}
\ham_n \!=\! \sqrt{2} \gamma_nB_1 \left(\alpha_{1}\!\ket{\hat{1}}\!\bra{\hat{1}}_e\!+\!\alpha_{0}\!\ket{\hat{0}}\!\bra{\hat{0}}_e\!+\!\alpha_{-1}\!\ket{\text{-}\hat{1}}\!\bra{\text{-}\hat{1}}_e\!\right)\hat{I}_x
\label{eq:RabiHam}
\end{equation}
Here $\alpha_{m_s}$ denote the enhancement factors in each manifold of the NV spin:
\begin{eqnarray}
		\alpha_{+1} \approx 1 &+&\frac{\gamma_e}{\gamma_n}\frac{A_\perp}{\Delta+\gamma_eB_z-\gamma_nB_z-Q}, \label{eq:enhancefactor1}\\
		\alpha_0  \approx  1&-&\frac{\gamma_e}{\gamma_n}\Bigl(\frac{A_\perp}{\Delta+\gamma_eB_z-\gamma_nB_z-Q} \nonumber \\
		&+&\frac{A_\perp}{\Delta-\gamma_eB_z-A_{\parallel}+\gamma_nB_z+Q} \Bigr), \label{eq:enhancefactor2}\\	
		\alpha_{-1} \approx  1&+&\frac{\gamma_e}{\gamma_n}\frac{A_\perp}{\Delta-\gamma_eB_z-A_{\parallel}+\gamma_nB_z+Q}, \label{eq:enhancefactor3}
		\end{eqnarray}	
		where we show expressions exact up to the first order in $\vartheta^\pm$ (see \cite{SMAPS} for the exact expressions).
The Hamiltonian $\ham_e$ can be neglected since electronic spin transitions are far off-resonance. 		

Thanks to the strong dependence of the enhancement factors on the transverse hyperfine coupling,  we can determine $A_\perp$ with high precision from measurement of the \Nit Rabi oscillations.

\textbf{Experiments --}
We used a home-built confocal microscope to measure the transverse hyperfine interaction of a single NV center in an electronic grade diamond sample  (Element 6, \Nit concentration $n_N<$~5 ppb, natural abundance of \carb).  The NV center is chosen to be free from close-by \carb.
	We worked at magnetic fields ($300$-$500$G) close to the excited state level anti-crossing so that during optical illumination at 532nm, polarization of the NV spin can be transferred to the nuclear spin by their strong hyperfine coupling in the excited state\cite{Jacques09}. As a result, a $1\mu s$ laser excitation polarizes the NV-\Nit system into the $\ket{0,1}$ state. 
	
\begin{figure}[t]\centering
\includegraphics[width= 0.45\textwidth]{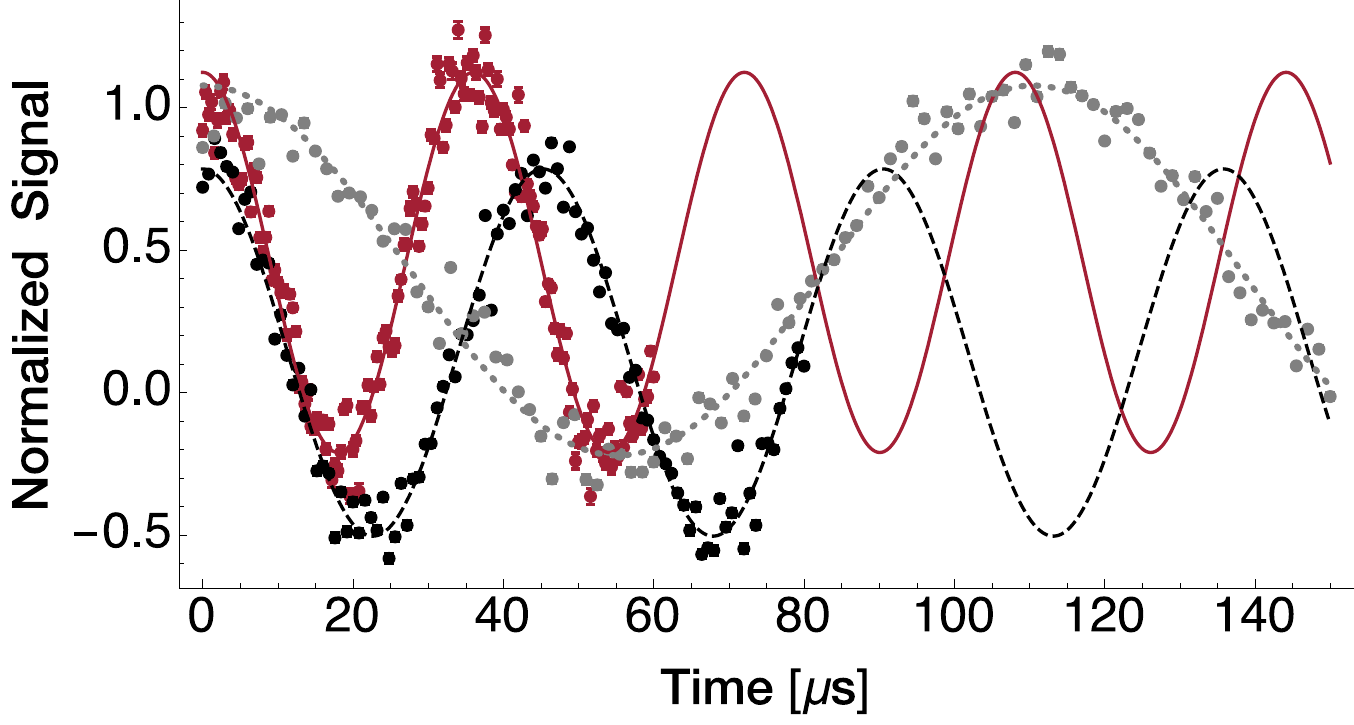}
\caption{\Nit Rabi oscillations at $B=450$G, $B_1 \approx 3.3$G in the three NV manifold (Red, solid line $m_s=0$. Black, dashed line, $m_s=-1$. Gray, dotted line $m_s=+1$). Here the dots are the experimental results, while the lines are fits to cosine oscillations. The different baseline of the $m_s=-1$ curve is due to small differences in the fluorescence emission of different nuclear manifolds\cite{Jacques09}.\label{fig:RabiOsc}}
\end{figure}
Then, the NV spin is prepared in the desired Zeeman state by a strong microwave (MW) pulse ($t_p\approx 50$ns)  before  coherently driving the nuclear spin by an RF field on resonance with the nuclear transition $\ket{m_s,1} \leftrightarrow \ket{m_s,0}$, for a duration $\tau$ (see Fig.~\ref{fig:experiment}). Finally, the nuclear spin state is detected by employing a MW selective pulse ($t_p\approx 700$ ns) that maps the nuclear spin state onto  the NV spin, which in turn can be read out  optically due to spin-dependent fluorescence emission intensity. The nuclear Rabi oscillations in Fig.~(\ref{fig:RabiOsc}) clearly show that for a fixed driving strength, the effective Rabi frequency is quite different in the three electronic spin manifolds.

\begin{figure}[t]\centering
\includegraphics[width= 0.48\textwidth]{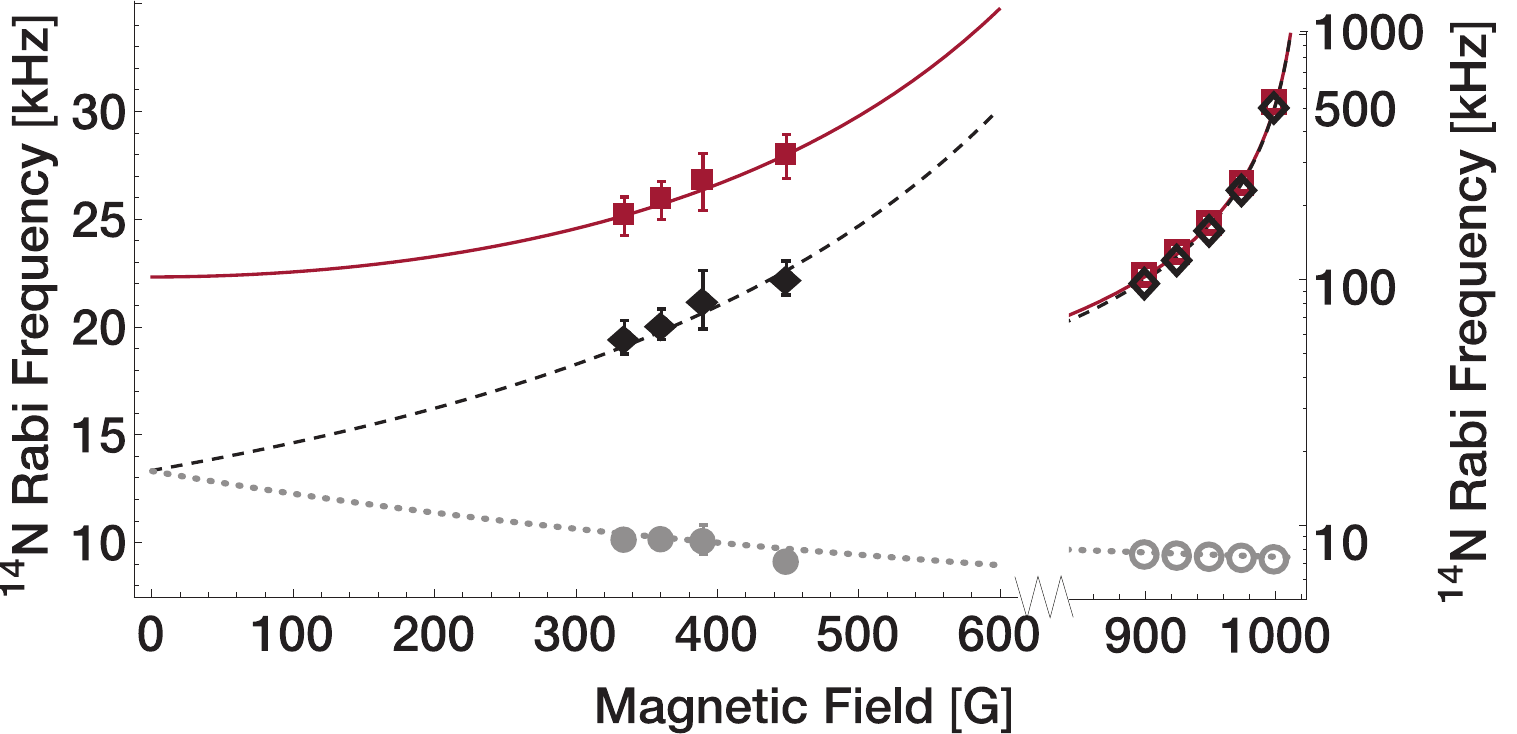}
\caption{\Nit Rabi Frequency in the three NV manifold (Red, solid line $m_s=0$. Black, dashed line, $m_s=-1$. Gray, dotted line $m_s=+1$) as a function of the magnetic field. Rabi frequency corresponds to $\frac{\gamma_n B_1}{\sqrt{2} \pi} \alpha_{m_s}$. The filled symbols correspond to the experimental data, which matches closely the theoretical prediction. The effective Rabi frequencies increase rapidly with the field, exceeding $1$ MHz when close to ground state level anti-crossing. The enhancement allows fast manipulation of the nuclear spin even when the bare Rabi field is only $B_1\approx 3.3$G. The theoretical prediction is confirmed by simulations (open symbols) of the spin dynamics.\label{fig:RabiVsBfield}}
\end{figure}

To confirm the expected dependence of the Rabi enhancement factors on the external magnetic field and the NV state, we measured the Rabi oscillations at the three electronic spin manifolds with varying magnetic field $B_z$. As shown in Fig.~(\ref{fig:RabiVsBfield}), the measured Rabi frequencies match well with the theoretical model. 
It is worth noting that contrary to the static pseudo-nuclear Zeeman effect\cite{AbragamBleaney}, there is a large enhancement ($\alpha_0\sim 16$, $\alpha_{\pm1}\approx -9$) even at zero field. Also, close to the ground state avoided crossing ($B\approx0.1$~T) the enhancement can become very large, exceeding 100. The validity of our approximation in this regime can be confirmed by numerical simulations\cite{SMAPS}.

	While these experiments could be used to extract $A_\perp$, this is not a practical method to obtain a good enough estimate.  The range of magnetic field is restricted by the need to be close to the excited state level anti-crossing, to achieve a good polarization of the nuclear spin. The number of acquired points is limited by the time it takes to change and properly align the external magnetic field. In addition,  there might be variations in the bare Rabi frequency in the three manifolds, because of different responses of the electronics used to drive the nuclear spins at the different frequencies.

In order to avoid these difficulties, we fixed the magnetic field to $509$G and instead linearly swept the amplitude of the RF driving ($B_1$).
	With this  procedure, we do not need an independent measure of the bare Rabi frequency in order to extract the transverse hyperfine coupling strength. The relative  RF amplitudes $B_1$ obtained when varying the driving strength can be measured at each nuclear resonance frequency by monitoring the RF voltage with an oscilloscope, confirming its linear dependence with applied power.  

\begin{figure}[t]
\includegraphics[width= 0.45\textwidth]{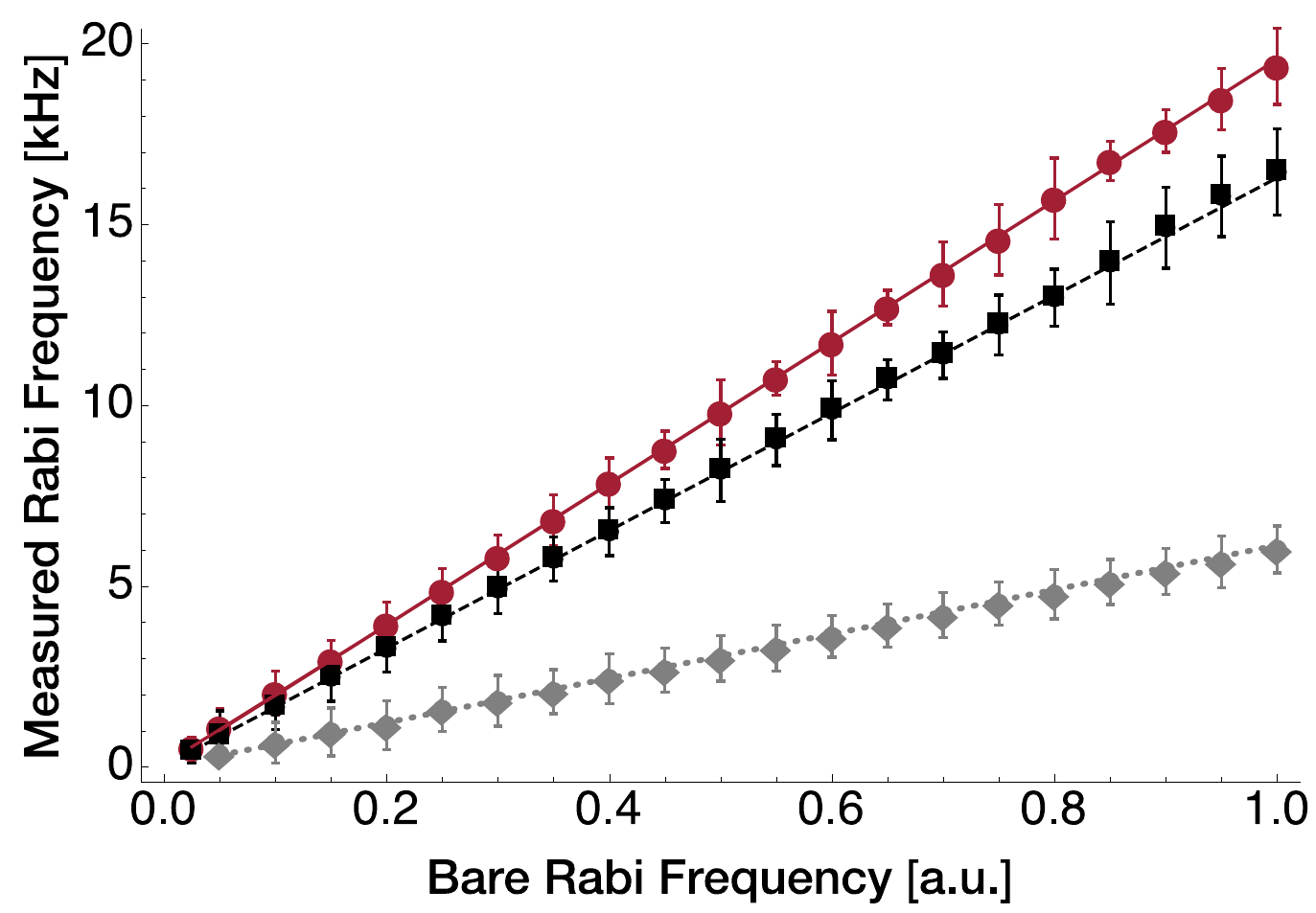}
\caption{Measured enhanced \Nit Rabi Frequency in the three NV manifold (Red, solid line $m_s=0$. Black, dashed line, $m_s=-1$. Gray, dotted line $m_s=+1$) as a function of the bare Rabi frequency at $B=509$G.\label{fig:FitPlot}}
\end{figure}

	We thus measure the effective nuclear Rabi frequency as a function of the normalized RF amplitude $B_1/|B_{1,max}|$ in all three electronic manifolds (Fig.~\ref{fig:FitPlot}). The measured Rabi frequency $\Omega_{m}$ is related to its on-resonance value by $\Omega_{m}=\sqrt{\Omega^2+\delta^2}$, where $\delta$ is the detuning from the nuclear spin resonance frequency. We incorporate this unknown, small detuning  in our model and fit the experimental data with the Rabi enhancement formulas (\ref{eq:enhancefactor1}-\ref{eq:enhancefactor3}). From the fit, we obtain an estimate of the transverse hyperfine coupling, $A_{\perp}=-2.62 \pm 0.05$~MHz, in good agreement with 
recently 
published values and with better precision than previously measured.

In order to achieve even better precision, we need to consider all the sources of uncertainty and errors. We find that small errors from imperfect MW $\pi$ pulses and nuclear polarization only contribute to a reduced fluorescent contrast, but do not affect the estimate of the Rabi frequency under our experimental condition. The detuning of the selective MW and RF pulses from resonance and uncertainty in $A_{\parallel}$ contributes only linearly to the uncertainty. All these minor errors and uncertainties affect very little the final uncertainty in the estimate of $A_\perp$\cite{SMAPS}.
The major source of error arises instead from the uncertainty in the measured Rabi frequency, which is limited by photon shot noise of the optical readout process. Therefore, the precision of the estimate could be improved with more averaging, at the expense of longer measurement time. Currently our total measurement time is limited by the stability of experimental setup, yielding $\delta A_{\perp} \sim 50$ kHz. Improving the stability of the setup by reducing thermal fluctuations and noise in the driving field (also using decoupling schemes\cite{Cai12,Aiello13}) or employing small ensembles or more efficient optical readout methods such as solid-immersion lenses\cite{Marseglia11} and charge-state sensing\cite{Shields15} could provide higher precision. 
Then, the limit would come from uncertainties in $\gamma_e$ and $\gamma_n$, with relative error of $10^{-4}$\cite{Doherty13,Neumann11t}, yielding an uncertainty in $A_\perp$ of a few hundred Hz\cite{SMAPS}.

\textbf{Conclusions --}
In conclusion, we observed enhanced nuclear Rabi oscillation in the NV-\Nit system due to level mixing between electronic and nuclear spin states. We harness the strong dependence of this enhancement on the transverse hyperfine coupling to determine its value with higher precision than previously published results. Theoretical analysis predicts an enhancement factor of almost 3 orders of magnitude when the magnetic field is close to the ground state level anti-crossing, promising fast manipulation of nuclear spin qubit at $\sim$~MHz rates, with only moderate driving strengths. More broadly, the method presented here can be applied to many other electron-nuclear hybrid spin systems to similarly characterize their interaction Hamiltonian with high precisions. Our results indicate that taking into account non-secular parts of a system's Hamiltonian is crucial to achieving faster and more accurate control of the quantum system.

\textbf{Acknowledgements --}
This work was supported in part by the U.S. Air Force Office of Scientific Research grant No. FA9550-12-1-0292 and by the U.S. Office for Naval Research grant No. N00014-14-1-0804.

\bibliographystyle{apsrev4-1}
\bibliography{Biblio}

\end{document}